\documentclass[a4paper]{article}
\begin {document}

\title {\bf QUANTUM NEWTON'S LAW }
\author{A.~Bouda\footnote{Electronic address: 
{\tt bouda\_a@yahoo.fr}} \ and T.~Djama\footnote{Electronic address: 
{\tt djama.toufik@caramail.com}}\\
Laboratoire de Physique Th\'eorique, Universit\'e de B\'eja\"\i a,\\ 
Route Targa Ouazemour, 06000 B\'eja\"\i a, Algeria\\}

\date{\today}

\maketitle

\begin{abstract}
\noindent
Using the quantum Hamilton-Jacobi equation within the framework of 
the equivalence postulate, we construct a Lagrangian of a quantum 
system in one dimension and derive a third order equation of motion 
representing a first integral of the quantum 
Newton's law. We then integrate this equation in the free particle 
case and compare our results to those of Floydian trajectories. 
Finally, we propose a quantum version of Jacobi's theorem. 
\end{abstract}

\vskip\baselineskip



\newpage

\vskip0.5\baselineskip
\noindent
{\bf 1\ \ Introduction }
\vskip0.5\baselineskip

Recently, Faraggi and Matone derived quantum mechanics from
an equivalence postulate which stipulates that all quantum 
systems can be connected by a coordinate
transformation \cite{FM1,FM1a,FM2}. They deduced that in one dimension, 
the physical solution of the stationary Schr\"odinger equation 
must have the form
\begin {equation}
\phi(x)= \left({\partial S_0 \over \partial x }\right)^{-1/2}
\left[\alpha\ \exp\left({i\over\hbar }S_0\right)
+\beta\ \exp\left(-{i\over\hbar }S_0\right)\right]\; ,
\end {equation} 
$\alpha$ and  $\beta$ being complex constants and $S_0$  
the reduced action. In contrast with Bohm's theory, this form of the
wave function indicates that the conjugate momentum, defined as
\begin {equation}
P={\partial S_0\over\partial x} \; ,
\end {equation}
never has a vanishing value both for bound and unbound states. 
In the case where the wave function $\phi$ is real up to a constant
phase factor, we have $|\alpha|=|\beta|$ but never $S_0=cte$.
Furthermore, by taking the derivative with respect to $x$ of the 
expression of $S_0$ given below, one can see that $P$ is always real 
even in classically forbidden regions. Within the framework of 
differential geometry, it is shown in Refs. \cite{FM1,FM1a,FM2} that the 
quantum stationary Hamilton-Jacobi equation (QSHJE), leading to 
the stationary Schr\"odinger equation in which the wave function 
is related to $S_0$ by Eq. (1), is
\begin {equation}
{1\over 2m} \left({\partial S_0 \over \partial x}\right)^2 + V(x)-E= 
{\hbar^2\over 4m}  \left[{3\over 2}\left( 
{\partial S_0 \over\partial x}\right)
^{- 2 }\left({\partial^2 S_0 \over \partial x^2}\right)^2-
\left( {\partial S_0 \over \partial x  }\right)^{- 1 }
\left({\partial^3 S_0 \over \partial x^3  }\right) \right] \; ,
\end {equation}
where $V(x)$ is the potential and $E$ the energy. In Eq. (3), the
left hand side reminds us of the classical Hamilton-Jacobi equation 
and the right one, proportional to $\hbar^2$ and called the quantum 
potential, describes the quantum effects. 
The higher dimension version of Eqs. (1) and (3) is obtained in 
Ref. \cite{BFM}. It is also obtained without appealing to 
differential geometry in Ref. \cite{Bouda}. The solution of 
Eq. (3), investigated also by Floyd \cite{F1,F2,F2a,F3} and 
Faraggi-Matone \cite{FM1,FM1a,FM2}, 
is given in Ref. \cite{Bouda} as
\begin {equation}
S_0=\hbar \ \arctan{\left({ \theta_1 + \nu \theta_2} \over 
{\mu \theta_1 +\theta_2}\right)} + \hbar \lambda \; ,
\end {equation}
where $\theta_1$ and $\theta_2$ represent two real independent 
solutions of the Schr\"odinger equation \  
$-\hbar^2 \phi''/ 2m + V \phi = E \phi$ \  and \  
$(\mu,\nu,\lambda)$ \ are real integration constants satisfying 
the condition $\mu\nu \not= 1$.

Trajectory representation of quantum mechanics, in which the
conjugate momentum is different from the mechanical one, was 
first introduced by Floyd \cite{F3,F4} who assumed that 
trajectories were obtained by using Jacobi's theorem
\begin {equation}
t-t_0={\partial S_0\over\partial E} \; .  
\end {equation}
\noindent
In classical mechanics, this theorem is a consequence of a
particular canonical transformation which is used in
Hamilton-Jacobi theory. Let us recall that the classical
Hamilton-Jacobi equation is a first order differential 
equation while the QSHJE is a third order one. 

In this letter, we propose a new procedure to determine the motion 
of any quantum system. In Sec. 2, we construct a Lagrangian from
which we derive in Sec. 3 the quantum law of motion. In Sec. 4, we 
integrate this law in the free particle case and in Sec. 5 we
compare our results to Floydian trajectories and propose a 
quantum version of Jacobi's theorem. 

\vskip0.5\baselineskip
\noindent
{\bf 2\ \ Construction of the Lagrangian }
\vskip0.5\baselineskip

First, let us remark that comparing to the usual classical reduced 
action in one dimension,  expression (4) for $S_0$ contains two 
additional integration constants $\mu$ and $\nu$ since $S_0$ depends 
also on the constant $E$ through the solutions $\theta_1$ and 
$\theta_2$ of the Schr\"odinger equation. This suggests that the 
fundamental law describing the quantum motion is a differential 
equation of fourth order since the Newton's classical law is a 
second order one. This means that the corresponding Lagrangian in 
the stationary case must be a function of $x$, $\dot{x}$, $\ddot{x}$
and maybe of $\dot{\ddot{x} }$ with a linear dependence.
However, it is not easy to construct from such a Lagrangian a 
formalism which leads to the well-known QSHJE. In order to surmount 
this difficulty, we propose a Lagrangian which is a function of $x$ 
and $\dot{x}$, as in classical mechanics, and for which we incorporate 
two integration constants playing the role of hidden variables.
We will later eliminate the indeterminacy introduced by these 
constants in the formalism by appealing to the QSHJE and its 
solution. The form of the Lagrangian that we postulate is
\begin {equation}
L(x,\dot{x},\mu,\nu)={1\over 2} m {\dot{x}}^2 f(x,\mu,\nu)-V(x) \; ,
\end {equation}
where $f(x,\mu,\nu)$ is a function which we will determine below.
The parameters $\mu$ and $\nu$ are the non-classical integration 
constants. As we will see, the function $f$, and therefore  $L$, 
depend also on the integration constant $E$ representing the 
energy of the system.

Now, let us show that the form (6) of the Lagrangian can be  
justified by appealing to the coordinate transformation, 
introduced by Faraggi and Matone \cite{FM2,FM3} and called 
a quantum transformation,
$$
x \to \hat {x} \; ,
$$
after which the QSHJE takes the classical form 
\begin {equation}
{1\over 2m}{\left(\partial \hat {S}_0(\hat{x})\over 
\partial \hat{x}\right)}^2+\hat{V}(\hat{x})=E \; .
\end {equation}
The called quantum coordinate $\hat{x}$ is given by 
\begin {equation}
\hat{x}=\int^{x}{{\partial S_0/ \partial x \over
 \sqrt{2m\left(E-V(x)\right)}}}\; dx \; .
\end {equation}
As shown by Faraggi and Matone \cite{FM2,FM3}, setting 
\begin {equation}
\hat{S}_0(\hat{x})=S_0(x),\ \ \ \ \ \ \ \ \ 
\hat{V}(\hat{x})=V(x) \; ,
\end {equation}
Eq. (7) takes the form
\begin {equation}
{1\over 2m}{\left(\partial S_0(x)\over \partial x\right)}^2
{\left({\partial x\over \partial \hat{x}}\right)}^2 +V(x)=E \; .
\end {equation}
The expression of the Hamiltonian can be obtained from (10)  
with the use of (2)
\begin {equation}
H={P^2\over 2m}{\left({\partial x\over
 \partial \hat{x}}\right)}^2 +V(x) \; .
\end {equation}
From (4) and (8), it is clear that $\hat{x}$ is a function  
depending on $x$ and on the parameters $(E, \mu, \nu)$. 
Therefore, as in classical mechanics, the velocity is 
given by the canonical equation
\begin {equation}
\dot{x}={\partial H\over \partial P}=
{P\over m}\ {\left(\partial x\over \partial \hat{x}\right)}^2 \; . 
\end {equation}
If we set
\begin {equation}
f(x,E,\mu,\nu)={\left(\partial \hat{x}
\over \partial x\right)}^2 \; ,
\end {equation}
the well-known relation \ $L=P\dot{x}-H$ \ leads to the form (6) 
of the Lagrangian. Using (10), we obtain the following expression  
for $ f(x,E,\mu,\nu)$
\begin {equation}
f(x,E,\mu,\nu)={1 \over 2m}{(\partial S_0/ \partial x )^2 
\over {E-V(x)}} \; .
\end {equation}
If we substitute $S_0$ with its expression (4), we effectively notice 
that $f$ depends on $x$, $E$, $\mu$ and $\nu$. Note that the coordinate 
$\hat{x}$ is real in classically allowed regions $(E>V)$ and purely 
imaginary in forbidden ones $(E<V)$. It follows that the function $f$ 
which we introduced in the Lagrangian is real positive in 
classically allowed regions but negative in forbidden ones. 
This means that in expression (6), the kinetic term in which the 
well-known quantum potential is absorbed is negative in classically 
forbidden regions although the velocity $\dot{x}$ is always real.

As observed by Faraggi and Matone \cite{FM2,FM3}, when we take the 
classical limit $\hbar \to 0$, the quantum coordinate $\hat{x}$ 
reduces to $x$. Then, using (13) the function $f$ 
goes to 1, leading in (6) to the classical form of the 
Lagrangian. We can therefore consider the Lagrangian (6) as a 
generalization of the classical one. 

Note also that if we define the conjugate momentum from (7) as 
$$
\hat {P}={\partial\hat{S_0}\over \partial \hat{x}}=
\left ({\partial S_0 \over\partial{x}}\right )
 \left ({\partial x \over \partial \hat{x}}\right ) \; ,
$$   
construct the Hamiltonian $\hat {H}$ and then the Lagrangian
$\hat {L}$, we get to the same expression (6) for the
Lagrangian by postulating the invariance of $L$ under the 
quantum transformation. However, in this construction, 
$\hat{P}$ is not real in classically forbidden regions.

\vskip0.5\baselineskip
\noindent
{\bf 3\ \ The quantum law of motion}
\vskip0.5\baselineskip

Using expression (6) of the Lagrangian, the least action principle 
leads to
\begin {equation}
mf\;{\ddot{x}} +\ {m \over 2}  {df \over dt} \dot{x}+
 {dV \over dx}=0 \; ,
\end {equation}
where we have used the fact that 
$\dot{x}{\left(\partial f/ \partial x\right)}={df / dt}$.
Integrating this last equation gives 
\begin {equation}
{1 \over 2}m\ {\dot{x}}^2 \ f \ +\ V\ =E \; ,
\end {equation}
which can be shown to be equivalent to (11) if we use (12) and (13). 
Note that the integration constant $E$ appearing in the
right hand side of (16) is already implicitly present in (15) 
through the function $f$. Substituting in (16) 
$f$ by its expression (14), we find 
\begin {equation}
{\partial S_0 \over \partial x}=
{2(E-V) \over \dot{x}} \; ,
\end {equation}
where we have eliminated one of the roots for 
$\partial S_0/\partial x$ since Eq. (12) indicates that $\dot{x}$ 
and $P = \partial S_0/\partial x$ have the same sign in 
classically allowed regions and are opposite in forbidden ones. 
Eq. (17), which is a consequence of both the
Lagrangian formulation and the QSHJE, will allow us 
to obtain a fundamental equation describing the quantum 
motion of any system. 

First, note that in the classical case, we have
$
{\partial S_{0}^{cl}/ \partial x}=m \dot {x}
$
and Eq. (17) reproduces the classical conservation equation 
$$
E={1\over 2} m{\dot {x}}^2+V(x) \; .
$$

Let us now derive the quantum conservation equation using 
the solution (4) of the QSHJE. Setting 
\begin {equation}
\phi_1=\mu \theta_1 +\theta_2 ,\ \ \ \ \ \
\phi_2=\theta_1 + \nu \theta_2 \; ,
\end {equation}
we have
$$
S_0={\hbar\;\arctan \left(\phi_2 \over \phi_1 \right)} \; ,
$$
and then 
\begin {equation}
 {\partial S_0 \over \partial x }=
{\hbar }{ {\phi_1 {\phi'_2}-
{\phi'_1}  \phi_2  } \over  
 {{\phi_1}^2+{\phi_2}^2 } }\; .
\end {equation}
From (17) and (19), we easily deduce 
\begin {equation}
\phi_1 {\phi'_2}-{\phi'_1}  \phi_2 
={2 \over \hbar }{E-V\over \dot {x}} \left ({\phi_1}^2+
{\phi_2}^2 \right ) \; .
\end {equation}
In order to eliminate the functions $\phi_1$ and $\phi_2$ 
and their derivatives, we differentiate this last equation with 
respect to $x$. Using the fact that the
derivative of the left hand side vanishes since it represents
the Wronskian of $ \phi_1$ and $ \phi_2 $ which are also
solutions of the Schr\"odinger equation, it follows that 
\begin {equation}
\phi_1 \phi'_1+\phi_2 \phi'_2 =
{1 \over 2}\left ( {1 \over E-V} {dV \over dx}+
 {\ddot {x} \over \dot{x}^2} \right) \left({\phi_1}^2+
{\phi_2}^2 \right ) \; .
\end {equation}
Now, differentiating (21) with respect to $x$ and using 
the fact that 
$$
{\phi''_1}=-{2 m \over {\hbar}^2}(E-V) {\phi_1}\; , \ \ \ \ \ \ \ \ 
{\phi''_2}=-{2 m \over {\hbar}^2} (E-V) {\phi_2} \; , 
$$
we find
\begin {eqnarray}
\left({{\phi'_1}}^2+{{\phi'_2}}^2 \right )-
\left ( {1 \over E-V}
 {dV \over dx}+ {\ddot {x} \over 
 \dot {x}^2} \right)
\left(\phi_1 \phi'_1 + \phi_2 \phi'_2 \right)
+\left [-{2m \over {\hbar}^2} (E-V) \right. \hskip5mm&& \nonumber\\
\left.-{1 \over 2} {1 \over (E-V)^2} \ \left({dV \over dx}\right)^2
- {1 \over 2(E-V)} {d^2V \over dx^2}
- \ {\dot {\ddot {x}} \over 2 \dot {x}^3}+
 {{\ddot {x}}^2 \over \dot {x}^4} \right ]\
 \left ({\phi_1}^2+{\phi_2}^2 \right )=0 \; .
\end {eqnarray}
If we solve the system constituted by (20) and (21)
with respect to $\phi'_1$ and $\phi'_2$, we can show that
\begin {eqnarray}
{{\phi'_1}}^2+{{\phi'_2}}^2 = \left [{4\over {\hbar}^2}
{(E-V)^2 \over {\dot{x}}^2}+ {1 
\over 4(E-V)^2}
{\left ({dV \over dx} \right)}^2\right. \hskip25mm&& \nonumber\\
\left.+{{\ddot{x}}^2 \over 4{\dot{x}}^4}+
 {1 \over 2(E-V)}{dV \over dx}
{\ddot{x} \over {\dot{x}}^2} \right]
\left({\phi_1}^2+{\phi_2}^2 \right) \; .
\end {eqnarray}
Now, if we substitute in (22) the quantities
$(\phi_1 \phi'_1+\phi_2 \phi'_2 )$
and
$({\phi'_1}^2+{\phi'_2}^2 )$ 
by their expressions (21) and (23), we find an equation in
which all the terms are proportional to 
$({\phi_1}^2+{\phi_2}^2)$. 
We then deduce 
\begin {eqnarray}
(E-V)^4-{m{\dot{x}}^2 \over 2}(E-V)^3+{{\hbar}^2 \over 8} 
{\left[{3 \over 2}
{\left({\ddot{x} \over \dot{x}}\right)}^2-{\dot{\ddot{x}} \over \dot{x}} 
\right]} (E-V)^2\hskip15mm&& \nonumber\\
-{{\hbar}^2\over 8}{\left[{\dot{x}}^2 
{d^2 V\over dx^2}+{\ddot{x}}{dV \over dx}
 \right]}(E-V)-{3{\hbar}^2\over 16}{\left[\dot{x}
{dV \over dx}\right]^2}=0 \; .
\end {eqnarray}
Because it depends on the integration constant $E$, this equation 
represents a first integral of the quantum Newton's law (FIQNL). 
It is a third order differential equation in $x$  
containing the first and second derivatives of the classical 
potential $V$ with respect to $x$. It follows that the solution 
$x(t,E,a,b,c)$ of (24) contains four integration constants which 
can be determined by the initial conditions
\begin {equation}
x(t_0)=x_0 \; , \ \ \ {\dot{x}(t_0)}={\dot{x}_0} \; , \ \ \ 
{\ddot{x}(t_0)}={\ddot{x}_0} \; , \ \ \
{\dot{\ddot{x}}(t_0)}={\dot{\ddot{x}}_0} \; . 
\end  {equation}
Of course, if we put $\hbar=0$, Eq. (24) reduces to the well-known 
first integral of the Newton's classical law 
$E= m {\dot{x}}^{2} / 2 + V(x)$. 
If we solve (24) with respect to $(E-V)$, then differentiate 
the obtained roots with respect to $x$, we will obtain 
the Quantum Newton's Law. It will be a fourth order differential 
equation in $x$ and will contain the first, second and  
third derivatives of $V$, while the classical law 
$
m{\ddot{x}}=-{ dV / dx}
$
is a second order differential equation and 
contains only the first derivative of $V(x)$. 

The second method to derive the FIQNL is to use expression (17) 
for ${\partial S_0 / \partial x}$ and compute the derivatives
\begin {equation}
{\partial ^2S_0 \over\partial x^2}=-{2\over \dot{x}}
{dV \over dx}-{2(E-V){\ddot{x}}
\over \dot{x}^3} \; ,
\end {equation}
and
\begin {equation}
{\partial ^3S_0 \over\partial x^3}=-{2\over \dot{x}}
{d^2 V\over dx^2}+{6(E-V){\ddot{x}}^2\over 
{\dot{x}}^5}-{2(E-V){\dot{\ddot{x}}} \over {\dot{x}}^4} 
+{4\ddot{x}\over {\dot{x}}^3} {dV \over dx} \; .
\end {equation}
Substituting these expressions in the QSHJE given by (3), 
we obtain the FIQNL, which is exactly the same as that 
written in (24). 

\vskip0.5\baselineskip
\noindent
{\bf 4\ \ The free particle case}
\vskip0.5\baselineskip

Let us examine the case of the free particle for which $V=0$. 
The FIQNL takes the form 
\begin {equation}
E^2-{m{\dot{x}}^2 \over 2} E+{{\hbar}^2\over 8} \left[{3\over 2} 
\left({\ddot{x}\over \dot{x}}\right)^2-{\dot{\ddot{x}} \over \dot{x}} 
\right]=0 \; .
\end {equation}
In order to solve this differential equation, let us introduce 
the variables  
\begin {equation}
U=\sqrt{2mE}\; x \; , 
\ \ \ \ \ \ \ \ q=\sqrt{2E \over m} \; t \; ,
\end {equation}
which have respectively the dimensions of an action and a distance.
In terms of these new variables, Eq. (28) takes the form
\begin {equation}
{1\over 2m}\left({dU \over dq} \right)^{2}-E={{\hbar}^2\over 4m} 
\left[{3\over 2}\left({dU \over dq}\right)^{-2} 
\left({d^2U \over dq^2}\right)^2-{\left({dU \over dq}\right)}^{-1}
\left( {d^3U \over dq^3}\right)\right] \; .
\end {equation}
This equation has exactly the same form as (3) when the potential has 
a vanishing value. This allows us to use the solution (4) to 
solve Eq. (30). However, if we set  
\begin {equation}
{\theta_3}={\mu}\theta_1+\theta_2 \; ,
\end {equation}
the solution (4) takes the form 
\begin {equation}
S_0=
{\hbar} \arctan{\left[(1-{\mu}{\nu}) {{\theta_1} \over
{\theta_3}}+\nu \right]}+{\hbar}{\lambda} \; .
\end {equation}
It follows that the solution of (30) can be written as 
\begin {equation}
U={\hbar} \arctan{\left[a {{\psi_1}\over{\psi_2}}+b \right]} 
+ U_0 \; ,
\end {equation}
where $a$, $b$ and $U_0$ are real integration constants satisfying 
the condition $a\not=0$ and $(\psi_1,\psi_2)$ is a set of two real 
independent solutions of the equation
$$
-{{\hbar}^2\over 2m} {d^2\psi \over dq^2}=E{\psi} \; .
$$
Choosing
$
{\psi_1}=\sin{\left({\sqrt{2mE}\; q }/ {\hbar}\right)}\
$
and
$
\ {\psi_2}=\cos{\left({\sqrt{2mE}\; q }/ {\hbar}\right)},
$
it follows that 
\begin {equation}
x(t)={\hbar\over\sqrt{2mE}} \arctan{\left[a \tan{\left({2Et 
\over\hbar}\right)}+b\right]}+x_0  \; .
\end {equation}
This relation represents the quantum time equation for the free
particle. It is clear that $x$ depends on four integration 
constants $(E,a,b,x_0)$. In the particular case $a=1$ and $b=0$, 
Eq. (34) reduces to the classical relation 
$$
x(t)= \sqrt{2E \over m}\;t+x_0 \; .
$$
This is compatible with the finding of Floyd \cite{F5,F6} who 
reproduced the classical results by attributing particular values 
to the non-classical integration constants. However, we do not have
the same trajectories. In the next section, we will explain this 
difference.

\vskip0.5\baselineskip
\noindent
{\bf 5\ \ Floydian trajectories and Jacobi's theorem }
\vskip0.5\baselineskip

In this section, we compare our results with those
obtained in Floydian trajectory formulation and propose 
a quantum version of Jacobi's theorem. 

Before going further, we would like to point out that Floyd's 
conjugate momentum \cite{F4} and ours, given in Eq. (12), are 
both different from the classical one \ $m\dot{x}$.  

Now, let us consider again the free particle case. Using expression 
(32) as a solution of the QSHJE and choosing  
$\theta_1= \sin{ ({\sqrt{2mE}\;x/ \hbar}  )}$
and 
$\theta_3= \cos { ({\sqrt {2mE}\;x/ \hbar}  )}$, 
we have
\begin {equation}
S_0=\hbar \ \arctan {\left [ a \tan {\left ({\sqrt {2mE} \; x 
\over \hbar} \right )} +b \right ]} +\hbar \lambda \; ,
\end {equation}
where we have used the notation $\ a=1-\mu \nu \ $ and $b=\nu$.
Floyd's trajectories are obtained by using Eq. (5) 
\begin {equation}
t-t_0=a\   
{\sqrt{2m/ E}\; x 
\over (a^2+b^2+1)+
\sigma \; \cos (2{\sqrt{2mE} \; x / \hbar}+\; \gamma)} \; ,
\end {equation}
where 
$$
\sigma=\sqrt{a^4+b^4+1+2a^2b^2+2b^2-2a^2}\;, \ \ \
\gamma=\arctan\left [{ 2ab \over a^2-b^2-1 }\right ] \; .
$$
First, the relation between $t$ and $x$ contains four
integration constants ($E$, $t_0$, $a$, $b$) as our result given in (34).
We remark also that the classical motion 
$ x= \sqrt {2E/m}\; (t-t_0)$
is obtained from Floyd's result, Eq. (36), or from ours, Eq. (34), 
by choosing the particular values $a=1$  and $b=0$ for the non-classical 
integration constants. Note that it is possible to rewrite the 
trigonometric term appearing in the denominator of Eq. (36) so as to 
show that this term will vanish for $a=1$ and $b=0$. Note also 
that the values of $a$  and $b$  with which we reproduce the 
classical results depend on the choice of the two independent 
solutions of the Schr\"odinger equation used in the calculation 
of the reduced action. On the other hand, in the classical 
limit ${\hbar \to 0}$, in both Floyd's approach \cite{F3,F5} 
and ours, a residual indeterminacy subsists except for the 
particular microstate for which $a=1$ and $b=0$. However, 
by averaging the classical limit of the expression of $t-t_0$ 
over one cycle of the trigonometric term, Floyd obtained the 
classical result 
\begin {equation}
<\lim_{\hbar \to 0} (t-t_0)>=\sqrt {m \over 2E} \; x \; .
\end {equation}
Using this procedure, we also obtain the same result after we 
express $t$ in terms of $x$ in (34). Another interesting 
question investigated by Floyd concerns microstates \cite{F2,F2a}. 
His conclusions, confirmed in Ref. \cite{Bouda}, 
indicate that trajectory representation manifests microstates not 
detected by the Schr\"odinger wave function for bound states. 
Obviously, our approach does not affect these conclusions. 

Despite these many common points, it is clear that our trajectories 
(34) are different from those of Floyd which can be written as 
in (36). This difference can be explained as follows.

Our first argument concerns the expression of the reduced action  as 
a function of time in the free particle case. In Floyd's approach, 
it is only in the classical limit $\hbar \to 0$ that we have  
\begin {equation}
\lim_{\hbar \to 0} S_0=2E(t-t_0) \; ,
\end {equation}
while in our approach, from Eq. (17) with $V=0$, we have 
$$
dx \; {\partial S_0 \over \partial x}=2E \; dt \; ,
$$
leading straightforwardly to  
\begin {equation}
S_0=2E(t-t_0) \; ,
\end {equation}
without taking the limit  $\hbar \to 0$. This result is 
in agreement with the fact that the reduced action is given by
\begin {equation}
S=S_0-Et = \int_{t_0}^{t}{L \;dt}=\int_{t_0}^{t}{E \;dt}=E(t-t_0) \; ,
\end {equation}
up to an additive constant term.

Our second argument concerns the use of relation (5). 
In our point of view, this  classical relation 
resulting in the particular canonical transformation used in 
the Hamilton-Jacobi theory, must be applied when we use the 
coordinate $\hat {x}$ with which the QSHJE takes the classical 
form given in (7). Then, we write 
\begin {equation}
t-t_0=\left[{\partial \hat{S}_0(\hat{x})
\over\partial E}\right]_{\hat{x}=cte} \; . 
\end {equation}
If we substitute in this 
last equation $\hat{S}_0(\hat{x})$ by  $S_0(x)$, we obtain Eq. (5).
After this substitution, the derivative with respect to $E$ 
does not keep $\hat{x}$ invariant as in (41) since Eqs. (13), 
(14) and (4) indicate that $\hat {x}$ is a function of $E$. 
This is the fundamental reason for which our trajectories 
differ from those of Floyd.

Let us remark that if we take the derivative of Eq. (41) with respect 
to $\hat {x}$ and then use (7), 
we obtain 
$$
{dt \over d \hat{x}}={\partial \over \partial \hat{x}}
 {\partial \hat{S}_0(\hat{x}) \over \partial E}=  
{\partial \over \partial E}
 {{\partial \hat {S}_0(\hat{x})}  \over \partial \hat {x}}=  
{\partial \over \partial E} \sqrt{2m(E-\hat{V}(\hat{x}))} \; .
$$
Because $\hat {V} (\hat {x})=V(x)$, we have
\begin {equation}
{dt \over dx}{\partial x \over \partial \hat{x} } =
{\sqrt{m\over 2(E-V(x))}} \; .
\end {equation}
Taking into account the expression of $\partial \hat{x}/\partial x$ 
which we deduce from (8), Eq. (42) leads to the fundamental relation (17). 
As explained at the end of Sec. (3), we can derive from (17) the 
FIQNL without using the Lagrangian formalism. It is therefore 
possible to obtain the FIQNL, as given in (24), straightforwardly 
from Jacobi's theorem, as written in (41), and the QSHJE.

We would like to add that Faraggi and Matone \cite{FM2} also derived 
an equation which is a first integral of the 
quantum analogue of Newton's law. This equation depends  
on the quantum potential and, like ours, is a third order differential
equation. However, Faraggi-Matone's derivation is based on  
relation (5), while our equation can be derived from (41). The two 
equations are thus different. Concerning Floyd's conjugate 
momentum \cite{F4} and  ours, they are also different since the 
Floyd's is obtained by using (5).

To conclude, we would like to emphasize that the fundamental
quantum law of motion (24) was obtained with two different 
methods from (17), which was itself obtained in two different 
contexts:

- a Lagrangian formulation by taking advantage of the fact that 
the solution of the QSHJE is known;

- a quantum version of Jacobi's theorem as written in (41).

\vskip\baselineskip
\noindent
{\bf REFERENCES}


\begin{enumerate}

\bibitem{FM1}
A. E. Faraggi and M. Matone, {\it Phys. Lett. B} 450, 34 (1999), 
hep-th/9705108.  

\bibitem{FM1a}
A. E. Faraggi and M. Matone, {\it Phys. Lett. B} 437, 369 (1998),
hep-th/9711028.

\bibitem{FM2}
A. E. Faraggi and M. Matone, {\it Int. J. Mod. Phys. A} 15, 1869 (2000),
hep-th/9809127.

\bibitem{BFM}
G. Bertoldi, A. E. Faraggi, and M. Matone, {\it Class. Quant. Grav.} 17, 3965
(2000), hep-th/9909201.

\bibitem{Bouda}
A. Bouda, {\it Found. Phys. Lett.} 14, 17 (2001), quant-ph/0004044.

\bibitem{F1}
E. R. Floyd, {\it Phys. Rev. D} 34, 3246 (1986).

\bibitem{F2}
E. R. Floyd, {\it Found. Phys. Lett.} 9, 489 (1996), quant-ph/9707051. 

\bibitem{F2a}
E. R. Floyd, {\it Phys. Lett. A} 214, 259 (1996).

\bibitem{F3}
E. R. Floyd, quant-ph/0009070.

\bibitem{F4}
E. R. Floyd, {\it Phys. Rev. D} 26, 1339 (1982). 

\bibitem{FM3}
A. E. Faraggi and M. Matone, {\it Phys. Lett. A} 249, 180 (1998),
hep-th/9801033.

\bibitem{F5}
E. R. Floyd, {\it Int. J. Mod. Phys. A} 15, 1363 (2000),
quant-ph/9907092.

\bibitem{F6}
E. R. Floyd, {\it Physics Essays} 5, 130 (1992). 

\end{enumerate}
\end {document}